\documentclass[aps,prb,twocolumn,groupedaddress,showpacs,floatfix]{revtex4-1}
\usepackage{graphicx}
\usepackage{dcolumn}
\usepackage{bm}
\usepackage{amsmath}

\begin{document}

\title{Unique nature of the lowest Landau level in finite graphene samples with
zigzag edges: Dirac electrons with mixed bulk-edge character}

\author{Igor Romanovsky}
\email{Igor.Romanovsky@gatech.edu}
\author{Constantine Yannouleas}
\email{Constantine.Yannouleas@physics.gatech.edu}
\author{Uzi Landman}
\email{Uzi.Landman@physics.gatech.edu}

\affiliation{School of Physics, Georgia Institute of Technology,
             Atlanta, Georgia 30332-0430}

\date{9 December 2010; Physical Review B, {\bf in press}}

\begin{abstract}
Dirac electrons in finite graphene samples with zigzag edges under high magnetic fields 
(in the regime of Landau-level formation) are investigated with regard to their 
bulk-type and edge-type character. We employ tight-binding calculations on finite 
graphene flakes (with various shapes) to determine the sublattice components of the 
electron density in conjunction with analytic expressions (via the parabolic cylinder 
functions) of the relativistic-electron spinors that solve the continuous Dirac-Weyl 
equation for a semi-infinite graphene plane. Away from the sample edge, the higher 
Landau levels are found to comprise exclusively electrons of bulk-type character (for 
both sublattices); near the sample edge, these electrons are described by edge-type 
states similar to those familiar from the theory of the integer quantum Hall effect for 
nonrelativistic electrons. In contrast, the lowest (zero) Landau level contains 
relativistic Dirac electrons of a mixed bulk-edge character without an analog in
the nonrelativistic case. It is shown that such mixed bulk-edge states maintain also in
the case of a square flake with combined zigzag and armchair edges. Implications for the 
many-body correlated-electron behavior (relating to the fractional quantum Hall effect) in 
finite graphene samples are discussed.     
\end{abstract}


\maketitle

\section{Introduction}

In the last few years, following the isolation of monolayer \cite{geim04} and the 
fabrication of epitaxial \cite{dehe04} graphene, the physical properties of graphene 
nanostructures (including elongated graphene ribbons and finite graphene samples and flakes)
have established themselves as a major research direction in condensed-matter physics. This 
development was propelled by theoretical predictions (see, e.g. Refs.\ 
\onlinecite{fuji96,dres96,waka99,son06,brfr06.1,able06,geim09}) that the electronic 
properties of graphene nanostructures are strongly affected by the presence and termination 
character (in particular, zigzag or armchair) of the graphene edges, suggesting an 
unparalleled versatility and potential for future nanoelectronics applications. 
Crucial to the realization of this perspective is the capability to characterize and
engineer edges with high purity and perfection, a need that has spurred an ever expanding 
experimental effort which has already yielded highly promising results.
\cite{camp09,dres09,geim10,lamb10,sump10,yang10,klit10}

In this context, a recent study \cite{yrl10} of ours addressed the influence of graphene 
edges upon the properties of correlated many-body fractional-quantum-Hall-effect (FQHE) 
states of Dirac electrons. The most recent experimental observation \cite{andr09,kim09,bao10} 
of such FQHE correlated states [in suspended monolayer \cite{andr09,kim09} and bilayer 
\cite{bao10} graphene samples under high magnetic fields $(B)$] 
has marked another milestone in demonstrating the potential of graphene not only for future 
technological applications, but also for studying novel fundamental physics behavior. In 
particular, in Ref.\ \onlinecite{yrl10} we showed that the Dirac-electron spinors in the 
lowest Landau level display a mixed bulk-edge character for graphene samples with zigzag 
edges, with the bulk component giving rise to the $\nu=1/3$ FQHE state (but with an 
attenuated strength), while the edge component is responsible for the insulating behavior 
observed \cite{andr09,kim09,ong08} at the Dirac neutrality point. 

Naturally, the complexity of the computational many-body treatment of the interelectron 
repulsion necessitated the use in Ref.\ \onlinecite{yrl10} of certain simplified 
assumptions, i.e., a circular shape for the graphene sample and an uninterrupted zigzag 
edge. In this paper, motivated by the widespread and ongoing experimental activity on 
perfect-graphene-edge engineering (see above), we present a systematic study of the 
properties of Dirac-electron states (with respect to both their bulk-type and edge-type 
character) forming the Landau levels in graphene nanostructures with more realistic shapes 
\cite{shapes,peet08,baha09,wimm10,eyang10,libi10} (namely, flakes with triangular, hexagonal, 
and square shapes). To this end, we utilize a combination of tight-binding calculations on 
graphene flakes with analytic expressions (via the parabolic cylinder functions 
\cite{wolfram,stegbook}) for the relativistic-electron spinors associated with the continuous 
Dirac-Weyl equation of a semi-infinite graphene plane. 
  
We demonstrate that, away from the graphene-flake edge, the higher Landau levels contain 
exclusively electrons of a bulk-type character (for both sublattices); near the graphene-flake 
edge, these electrons are described by edge-type states reminiscent of those familiar from the 
theory of the integer quantum Hall effect for nonrelativistic electrons. \cite{halp82} In 
contrast, the lowest (zero) Landau level of the graphene flakes contains relativistic Dirac 
electrons of a mixed bulk-edge character without an analog in the nonrelativistic case. It is 
shown that such mixed bulk-edge states maintain also in the case of a square flake with 
combined zigzag and armchair edges. 

The paper is organized as follows:

Sec. \ref{secmeth} is devoted to the description of the methodologies employed. Specifically, 
Sec. \ref{secdw} derives the analytic expressions for the Dirac-Weyl spinors in the case of a 
semi-infinite graphene plane with zigzag edge termination; the solutions for both the
$K$ (Sec. \ref{seck}) and $K^\prime$ (Sec. \ref{seckpr}) graphene valleys are given. 
An outline of the tight-binding approach used here is given in Sec. \ref{sectb}.

Our tight-binding results at high magnetic field (concerning the electron-density components of
the two graphene sublattices and their interpretation through comparison with the 
continuum-model Dirac-Weyl spinors) are presented in Sec. \ref{sectri} for triangular flakes, 
Sec. \ref{sechex} for hexagonal flakes, and Sec. \ref{secsqu} for square flakes.

Finally, Sec. \ref{secsum} offers a Summary.

\section{Methodology}
\label{secmeth}

\subsection{Solutions of the Dirac-Weyl equation for a semi-infinite graphene plane}
\label{secdw}

%*********************** begin figure 1 **************
\begin{figure}[t]
\centering\includegraphics[width=7.8cm]{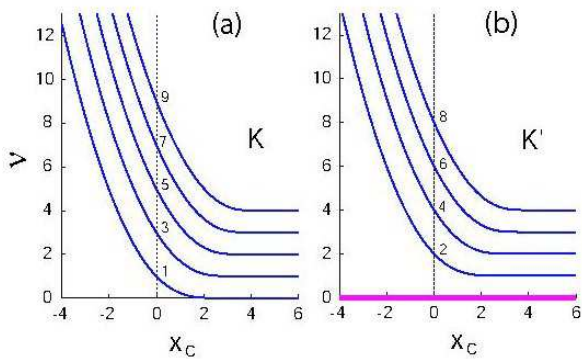}
\caption{%%%%
(Color online) The eigenenergies (specifically the index $\nu$) that solve the transcendental
equation as a function of $x_c$ (a) for the $K$ valley [see Eq.\ (\ref{eqtran})] and
(b) for the $K^\prime$ valley [see Eq.\ (\ref{eqtranpr})]. The almost flat segments of the 
curves correspond to Landau levels with energies $\varepsilon \approx \sqrt{2n}$, $n=0$, 1, 
2, $\ldots$; the approximate symbol $\approx$ signifies that $\nu$ does not take integer 
values, but comes extremely close to them. The rising-in-energy branches correspond to 
double-edge states (i.e., states of edge character on both the A and B sublattices); they 
cross the vertical axis at $x_c=0$ for (a) odd integer values and (b) even integer values.
Note that the $K$ valley (a) exhibits a dispersive (varying with $x_c=q_yl_B$) quasiflat band
with $\nu \approx 0$, while the $K^\prime$ valley (b) exhibits a dispersionless flat band with
$\nu = 0$ [see text and thick line (magenta color online)]; this is the only case when $\nu$ 
takes an integer ($\nu=0$) value.
}
\label{ene}% must come after the caption
\end{figure}
%*********************** end figure 1 **************

\subsubsection{$K$ valley}
\label{seck}

For a semi-infinite graphene plane under a perpendicular magnetic field $B$ (with the 
graphene plane extending for $0 \leq x < \infty$ and exhibiting a zigzag edge along the 
$y$ axis at $x=0$), the Dirac-electron wave function corresponding to the $K$ valley 
can be written as a two-component spinor (the zigzag boundary condition does not couple 
the two graphene valleys) 
\begin{equation}
\psi(x,y)=\frac{e^{i q_y y}}{\sqrt{2}} \left( 
\begin{array}{c}
\chi_A (x) \nonumber \\
\chi_B (x)
\end{array} 
\right).
\label{dspin}
\end{equation}
In Eq.\ (\ref{dspin}), $q_y=k_y-K_y$, with $k_y$, $K_y$ being the linear momenta of the 
electron and the $K$ valley along the $y$ direction; the magnetic length 
$l_B=\sqrt{\hbar c/eB}$. 

With the introduction of reduced (dimensionless) variables $x/l_B \rightarrow x$ and
$x_c = q_y l_B$, the continuous Dirac-Weyl equation coupling the $\chi_A (x)$ and
$\chi_B (x)$ components is given by
\begin{subequations}
%\label{allequations} % notice location
\begin{eqnarray} 
\frac{d}{dx} \chi_B + (x-x_c)  \chi_B &=& \varepsilon  \chi_A \label{dw1}\\
\frac{d}{dx} \chi_A - (x-x_c)  \chi_A &=& - \varepsilon  \chi_B \label{dw2},
\end{eqnarray}
\label{dw}
\end{subequations}  
\noindent
where the reduced energy $\varepsilon=E/(\hbar v_F/l_B)$, with $v_F$ being the Fermi 
velocity of graphene. 

%*********************** begin figure 2 **************
\begin{figure}[t]
\centering\includegraphics[width=7.5cm]{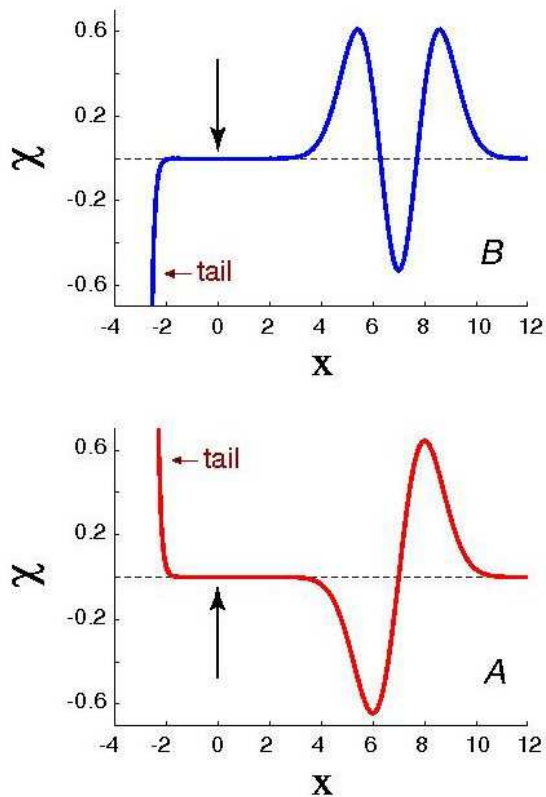}
\caption{%%%%
A bulk-bulk state in the third ($n=2$) LL: The $\chi_B$ (top frame) and $\chi_A$ (bottom 
frame) Dirac-spinor components are displayed for $x_c=7$ and $\nu=2+9.229 \times 10^{-18}$ 
[see the transcendental Eq.\ (\ref{eqtran})]. This state lies well inside the quasiflat 
segment of the third LL curve in Fig.\ \ref{ene}(a). The vertical arrows mark the position 
of the boundary at $x=0$. In the range $-\infty< x_c < 0$, note the development of 
exponentially growing tails, associated with the fact that the value of $\nu$ above is very 
close, but not equal, to an integer (here 2). Apart from the tails, both orbitals portrayed
here are very close to the eigenfunctions of a 1D harmonic oscillator centered at $x_c$; see 
Eq.\ (\ref{herm}). 
}
\label{chibaLL2}% must come after the caption
\end{figure}
%*********************** end figure 2 **************

%*********************** begin figure 3 **************
\begin{figure}[t]
\centering\includegraphics[width=7.5cm]{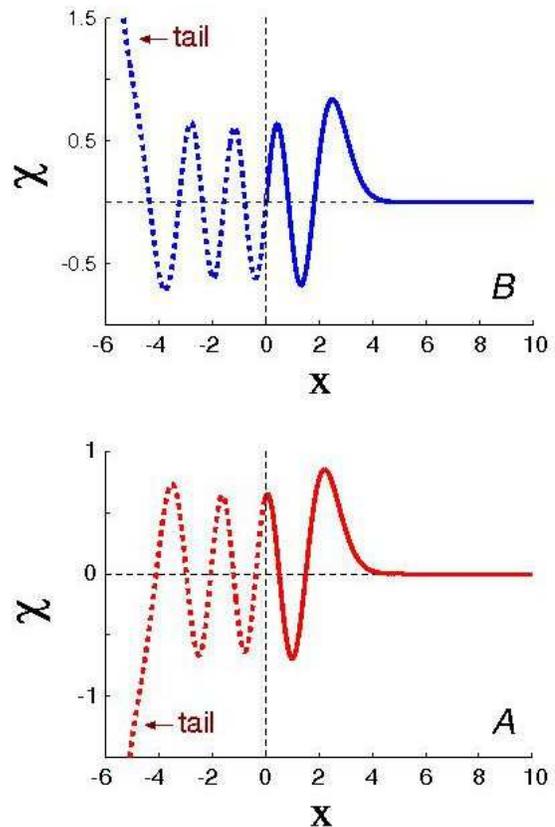}
\caption{%%%%
A double-edge state associated with the third ($n=2$) LL: 
The $\chi_B$ (top frame) and $\chi_A$ 
(bottom frame) Dirac-spinor components are displayed for $x_c=-1$ and $\nu=7.5266$ [see the 
transcendental Eq.\ (\ref{eqtran})]. This state lies on the rising branch extending out 
from the flat segment of the third LL curve in Fig.\ \ref{ene}(a). The vertical dashed lines
mark the position of the physical boundary at $x=0$. In the range $-\infty< x_c < 0$, note 
the development of exponentially growing tails, associated with the fact that the value of 
$\nu$ above is not equal to an integer. In the range $-\infty < x < +\infty$, the number of 
zeros associated with the B component is $\lceil \nu \rceil=8$; for the A component 
it is $\lceil \nu-1 \rceil=7$. The physically relevant range $0 \leq x < +\infty$ contains
only two zeros for both the cases of the B and A spinor components.
}
\label{chibaLL22}% must come after the caption
\end{figure}
%*********************** end figure 3 **************

In general, and prior to invoking any boundary conditions (that is considering
the complete graphene sheet for $-\infty < x < \infty$), the solutions of the
system of coupled equations in Eq.\ (\ref{dw}) fall into two classes, i.e., for
$\varepsilon \neq 0$ and $\varepsilon = 0$.

{\it Solutions for $\varepsilon \neq 0$.\/} 
In this case, one can multiply both sides of Eq.\ (\ref{dw2}) with
$\varepsilon$, and then use Eq.\ (\ref{dw1}) to eliminate $\chi_A$. The result is
the following second-order equation for $\chi_B$:
\begin{equation}
\frac{d^2}{d\xi^2} \chi_B(\xi)+ 
\left( \nu+\frac{1}{2}-\frac{1}{4}\xi^2 \right) \chi_B(\xi)=0,
\label{weber}
\end{equation}  
where 
\begin{equation}
\xi=\sqrt{2}(x-x_c) \text{~~and~~~}  \nu=\varepsilon^2/2.
\label{xinu}
\end{equation}

Eq.\ (\ref{weber}) has the standard form of a Weber differential equation, and thus its 
solutions coincide with the parabolic cylinder functions, \cite{wolfram,stegbook,gusi08} i.e.,
\begin{equation}
\chi_B(\xi) = C_\nu D_\nu(\xi),
\label{chib}
\end{equation}
where $C_\nu$ is a normalization constant.

%*********************** begin figure 4 **************
\begin{figure}[t]
\centering\includegraphics[width=7.5cm]{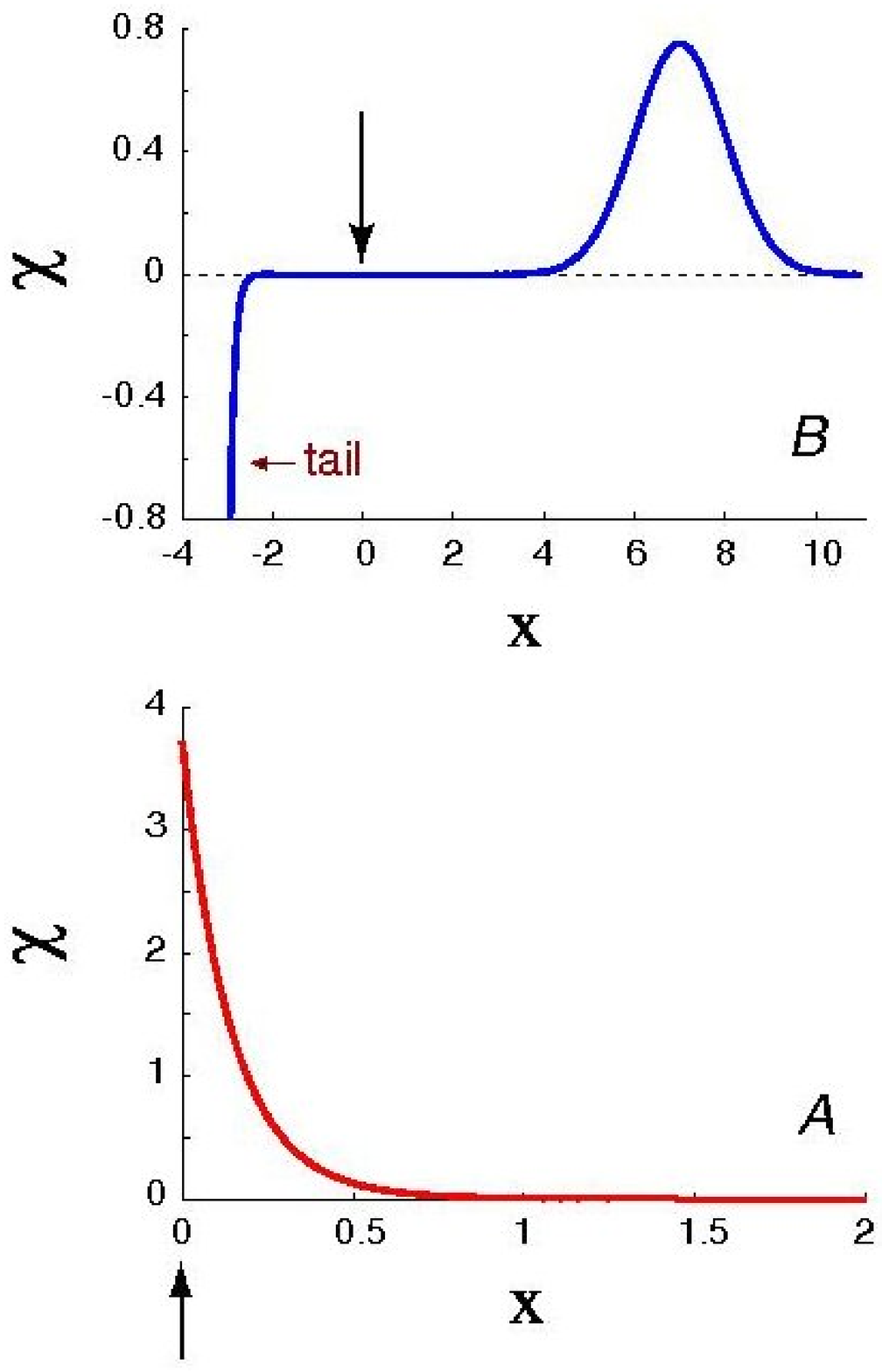}
\caption{%%%%
A bulk-edge state in the LLL ($n=0$): The $\chi_B$ (top frame) and $\chi_A$ (bottom frame) 
Dirac-spinor components are displayed for $x_c=7$ and $\nu=2.049 \times 10^{-21}$ [see the 
transcendental Eq.\ (\ref{eqtran})]. This state lies well inside the quasiflat segment of the
LLL curve in Fig.\ \ref{ene}(a). The vertical arrows mark the position of the boundary at 
$x=0$. In the top frame, note the development (in the range $-\infty< x_c < 0$) of an 
exponentially growing tail, associated with the fact that the value of $\nu$ above is very 
close, but not equal, to an integer (here 0). Apart from the tail, the orbital portrayed 
in the top frame is very close to the ground-state eigenfunction of a 1D harmonic oscillator 
centered at $x_c$; see Eq.\ (\ref{herm}).
}
\label{chibaLLL}% must come after the caption
\end{figure}
%*********************** end figure 4 **************

Using Eq.\ (\ref{dw1}), the recurrence relation
\begin{equation}
\frac{d}{d\xi}D_\nu(\xi) +\frac{1}{2} \xi D_\nu(\xi) - \nu D_{\nu-1}(\xi)=0,
\label{pcfrec}
\end{equation}
and Eq.\ (\ref{xinu}), the corresponding A component is given by:
\begin{equation}
\chi_A(\xi)= C_\nu \sqrt{\nu} D_{\nu-1}(\xi).
\label{chia}
\end{equation}

When $\nu$ is a nonnegative integer, $n \geq 0$, the parabolic cylinder functions reduce
to the familiar wave functions of the one-dimensional unconfined harmonic oscillator,
\begin{equation}
D_n(\xi)=2^{-n/2} e^{-\xi^2/4} H_n \left( \frac{\xi}{\sqrt{2}} \right),
\label{herm}
\end{equation}
where $H_n$ are Hermite polynomials. The wave functions in Eq.\ (\ref{herm}) have the 
property $D_n(\pm \infty) =0$, appropriate for an unconfined harmonic oscillator; they also 
exhibit $n$ zeros. When $\nu \neq n > 0$, $D_\nu(+\infty)=0$, and for $\xi <0$ 
an additional zero (compared to the case of $\nu=n$) develops, through which the parabolic 
cylinder function crosses the $x$-axis and then develops an exponentially growing tail;
an example for a state with (energy) $\nu=2+9.229 \times 10^{-18}$ and $x_c=7$ [see Fig.\
\ref{ene}(a)] is given in Fig.\ \ref{chibaLL2}. Specifically the number of zeros of 
$D_{\nu}(\xi)$ (with $\nu > 0$) is given by the ceiling function \cite{note1} 
$\lceil \nu \rceil$ (this includes the case when $\nu$ is a positive integer $n$); for 
$\nu \leq 0$, $D_{\nu}(\xi)$ has no zeros.

In the case of a semi-infinite graphene sheet with zigzag edges (extending for
$0 \leq x < \infty$), one can require that
the additional zero for the $\chi_B$ spinor component coincides with the origin of axes 
($x=0$); this provides the following transcendental equation for determining the energy 
levels of the Dirac electrons (remember that $\nu=\varepsilon^2/2$):
\begin{equation}
D_\nu(-\sqrt{2} x_c)=0.
\label{eqtran}
\end{equation}  

The single-particle energies $\varepsilon$ [which are solutions of Eq.\ (\ref{eqtran})] as a 
function of $x_c$ are displayed in Fig.\ \ref{ene}(a). One sees that Landau levels (with
energy $\varepsilon \approx \sqrt{2n}$, $n=0$, 1, 2, $\ldots$) are formed when the centroid
$x_c$ of the orbitals is far away from the physical edge; the approximate symbol $\approx$ 
signifies that $\nu$ does not take integer values, but comes extremely close to them. An 
illustrative case of the corresponding Dirac-spinor orbitals $\chi_B$ and $\chi_A$ are 
portrayed in Fig.\ \ref{chibaLL2}. In the domain $-\infty < x < 0$, a tail develops due to the
fact that the index $\nu$ is not an integer. In the physically relevant domain 
$0 \leq x < +\infty$, the two components are bulk-like and very similar to the familiar wave 
functions of a 1D unconstrained harmonic oscillator. Similar orbitals (differing only in the 
number of zeros) apply for all Landau levels with $\nu \approx n \geq 1$. 

For positive values of $x_c$ near the boundary, and also for negative values of $x_c$, 
double-edge states are formed reminiscent of the single-edge states familiar from the theory 
of the integer quantum Hall effect. \cite{halp82} The corresponding orbitals
for an illustrative case (with $x_c=-1$ and $\nu=7.5266$) are displayed in Fig.\ 
\ref{chibaLL22}. Again, one sees the development of a tail in the domain $-\infty < x < 0$, 
due to the fact that the index $\nu$ is not an integer.

The case of large and positive $x_c$ in the LLL ($n=0$) is special and of particular
significance regarding the strongly correlated Dirac-electron states in finite graphene 
samples under high magnetic field.\cite{yrl10} Indeed in this case, the Dirac spinor 
contains orbitals of both bulk and edge character. An illustrative case (with $x_c=7$ and 
$\nu=2.049 \times 10^{-21}$ is portayed in Fig.\ \ref{chibaLLL}. One sees that the B 
component is bulk-like and similar to the ground-state of an 1D harmonic oscillator in the 
physically relevant domain $0 \leq x < +\infty$. However, in the same domain, the A 
component is clearly edge-like. 

Further understanding of this LLL behavior can be achieved through the observation that
for $x_c >> 0$ the LLL $\chi_A$ component can be approximated by
\begin{eqnarray}
\chi_A^{\text{LLL,app}} (x) &\approx& \widetilde{C} D_{-1}(\xi) \nonumber \\ 
&=& \widetilde{C} e^{\frac{1}{2}(x-x_c)^2} \sqrt{\frac{\pi}{2}} \text{erfc}(x-x_c).
\label{dm1}
\end{eqnarray}
Taking into consideration that $\text{erfc}(-x_c) \rightarrow 2$ for (large) $x_c >> 1$,
and keeping the lowest order in $x/x_c$ in the exponent, one can determine the normalization
constant $\widetilde{C}$. The final simplified expression is:
\begin{equation}
\chi_A^{\text{LLL,app}} (x) = \sqrt{2 x_c} e^{-x x_c}.
\label{dm12}
\end{equation}

Eq.\ (\ref{dm12}) has the form of an exponential function decaying inside the graphene sheet.
This form agrees very well with the full solution of $\chi_A$ in Fig.\ \ref{chibaLLL} [see 
Eq.\ (\ref{chia}) with $x_c=7$ and $\nu=2.049 \times 10^{-21}$]. We note that the surface 
character of this edge-like LLL A-component becomes more pronounced (i.e., it exhibits a 
narrower $1/x_c$ width) the larger the (positive) value of the centroid $x_c$.  

%*********************** begin figure 5 **************
\begin{figure}[t]
\centering\includegraphics[width=7.5cm]{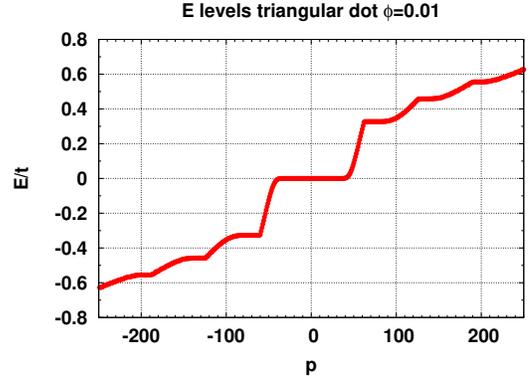}
\caption{%%%%
The TB energies at high magnetic field ($\phi=0.01$) for a triangular flake with zigzag 
edges comprising $N=6397$ carbon atoms. The $n=0$, $\pm 1$, $\pm 2$, and $\pm 3$ Landau levels 
correspond to the flat segments of the curve. The double-edge states correspond to the 
rising-in-energy branches between the flat segments. The integer index $p$ counts the TB states
(negative $p$ values correspond to negative energies). 
}
\label{enetb}% must come after the caption
\end{figure}
%*********************** end figure 5 **************

{\it Solutions for $\varepsilon = 0$.\/} In this case, the two equations in Eq.\ (\ref{dw}) 
decouple yielding the two solutions
\begin{subequations}
%\label{allequations23} % notice location
\begin{eqnarray}
  \chi_B(\xi) &=& C_B e^{-\xi^2/4} \label{e011}\\
  \chi_A(\xi) &=& 0 \label{e012},
\end{eqnarray}
\label{e01}
\end{subequations}
\noindent
and 
\begin{subequations}
%\label{allequations33} % notice location
\begin{eqnarray}
  \chi_B(\xi) &=& 0 \label{e021}\\
  \chi_A(\xi) &=& C_A e^{\xi^2/4} \label{e022}.
\end{eqnarray}
\label{e02}
\end{subequations}
\noindent

For the nontrivial case ($C_A$ or $C_B \neq 0$), neither of these two solutions satisfy the
boundary conditions $\chi_B(-\sqrt{2}x_c)=0$ and $\chi_B(+\infty)=\chi_A(+\infty)=0$. Thus
there is no dispersionless solution with $\varepsilon=0$ associated with the $K$ valley. 
However, as we will see below, such $\varepsilon=0$ solutions exist for the $K^\prime$ valley.

\subsubsection{$K^\prime$ valley}
\label{seckpr}

The continuous Dirac-Weyl equation coupling the $\chi_A^\prime (x)$ and 
$\chi_B^\prime (x)$ components in graphene's $K^\prime$ valley is given by
\begin{subequations}
%\label{allequationspr} % notice location
\begin{eqnarray} 
\frac{d}{dx} \chi_B^\prime - (x-x_c)  \chi_B^\prime &=& - \varepsilon  
(-\chi_A^\prime) \label{dw1pr},\\
\frac{d}{dx} (-\chi_A^\prime) + (x-x_c)  (-\chi_A^\prime) &=& \varepsilon  
\chi_B^\prime. \label{dw2pr}
\end{eqnarray}
\label{dwpr}
\end{subequations}  
\noindent

We note that Eq.\ (\ref{dwpr}) has the same form as Eq.\ (\ref{dw}) with the substitution 
$\chi_A \leftrightarrow \chi_B^\prime$ and $\chi_B \leftrightarrow - \chi_A^\prime$. 
As a result, for $\varepsilon \neq 0$, one has the following solutions for the Dirac
spinor in the $K^\prime$ valley
\begin{subequations}
%\label{allequationspr2} % notice location
\begin{eqnarray}
\chi_B^\prime(\xi) &=& C_\nu \sqrt{\nu} D_{\nu-1}(\xi) \label{chibpr} \\
\chi_A^\prime(\xi) &=& - C_\nu D_\nu(\xi), \label{chiapr}
\end{eqnarray}
\label{chibapr}
\end{subequations}
with the index $\nu=\varepsilon^2/2$ as was the case in the $K$ valley.

The transcendental equation in the $K^\prime$ valley for the indices $\nu$ (or energies 
$\varepsilon$) as a function of $x_c$ is written as
\begin{equation}
D_{\nu-1}(-\sqrt{2}x_c)=0.
\label{eqtranpr}
\end{equation}

The solutions of Eq.\ (\ref{eqtranpr}) as a function of $x_c$ are plotted in Fig.\ 
\ref{ene}(b). Note that the index $\nu > 1$ in this case [$D_{\nu-1}(\xi)$ has no
zeros for $\nu \leq 1$]. This contrasts with the case of the $K$ valley shown in Fig.\ 
\ref{ene}(a), where $\nu >0$.

For $\varepsilon=0$, the two equations in Eq.\ (\ref{dwpr}) decouple, and there is a
physically valid solution
\begin{subequations}
\begin{eqnarray}
  \chi_A^\prime(\xi) &=& -C^\prime_A e^{-\xi^2/4} \label{e011pr}\\
  \chi_B^\prime(\xi) &=& 0 \label{e012pr}.
\end{eqnarray}
\label{e01pr}
\end{subequations}
\noindent
Assuming a relation $\nu=\varepsilon^2/2$, this dispersionless band of edge states can be 
associated with an index $\nu=0$, and it is denoted by a thick dashed line in Fig.\ 
\ref{ene}(b). This band maintains also for zero-magnetic field, since
\begin{equation}
\lim_{B \to 0}  e^{-\xi^2/4} \propto e^{\tilde{x} q_y},
\label{limit}
\end{equation}
which represents an edge state for $q_y < 0$; the tilded $\tilde{x}$ denotes the $x$ position
in the original dimensions of length (before the introduction of the reduced variable 
$x=\tilde{x}/l_B$; see Ref.\ \onlinecite{brfr06.1}).

%*********************** begin figure 6 **************
\begin{figure}[t]
\centering\includegraphics[width=7.5cm]{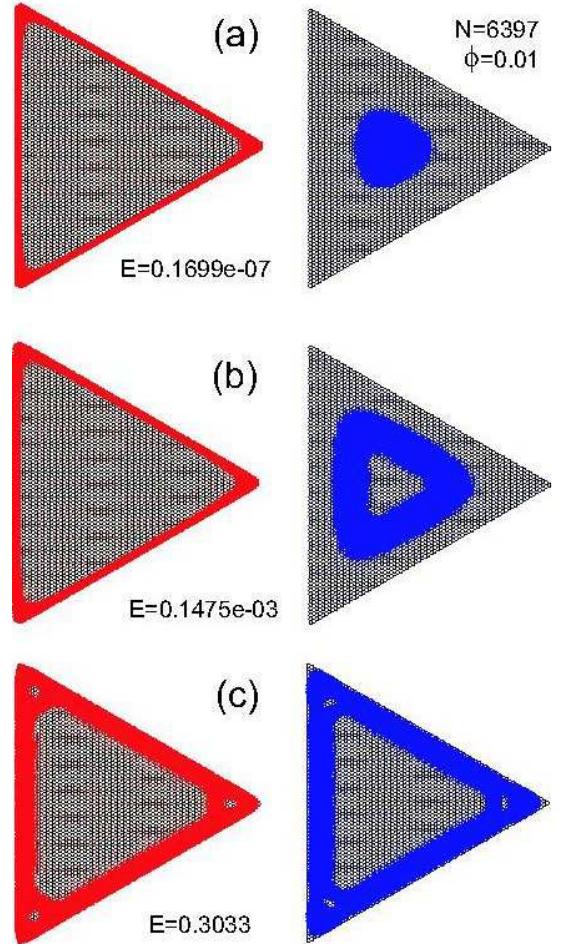}
\caption{%%%%
Examples of TB electron densities at high magnetic field ($\phi=0.01$) for the A 
(left) and B (right) sublattices associated with: (a)$+$(b) mixed bulk-edge dispersive states 
in the $n=0$ Landau level (whose TB energies reside on the flat step at $E \sim 0$ in Fig.\ 
\ref{enetb}). (c) double-edge states situated on the rising-in-energy branch between the $n=0$ 
and $n=1$ Landau levels.
Energies in units of the hopping coupling parameter $t$. 
}
\label{abtbLLL}% must come after the caption
\end{figure}
%*********************** end figure 6 **************

\subsection{Tight-binding approach for finite graphene flakes}
\label{sectb}

In the tight-binding (TB) calculations, we use the hamiltonian
\begin{equation}
H_{\text{TB}}= - \sum_{<i,j>} t_{ij} c^\dagger_i c_j + h.c.,
\label{htb}
\end{equation}
with $< >$ indicating summation over the nearest-neighbor \cite{note22}
sites $i,j$. The hopping
matrix element 
\begin{equation}
t_{ij}=t \exp \left( \frac{ie}{\hbar}  \int_{{\bf r}_i}^{{\bf r}_j} 
d{\bf s} \cdot {\bf A} ({\bf r}) \right), 
\label{tpei}
\end{equation}
where $t=2.7$ eV, ${\bf r}_i$ and ${\bf r}_j$ are the positions of the carbon atoms
$i$ and $j$, respectively, and  ${\bf A}$ is the vector potential associated with the
applied perpendicular magnetic field $B$. 

The calculations were carried out for two shapes that support a zigzag edge on all 
sides of the graphene flake, that is, equilateral triangles and regular hexagons, as well as
for a square shape which exhibits both zigzag and armchair edges. The number of carbon atoms 
considered is $N=6397$ for the triangular flakes, $N=6144$ for the hexagonal ones, and $N=2074$
for the square one. The diagonalization of the TB hamiltonian [Eq.\ (\ref{htb})] is 
implemented with the use of the sparse-matrix solver ARPACK.\cite{arpack}  

\section{Results of tight-binding calculations and their interpretation}

\subsection{Trigonal graphene flakes}
\label{sectri}

%*********************** begin figure 7 **************
\begin{figure}[t]
\centering\includegraphics[width=7.5cm]{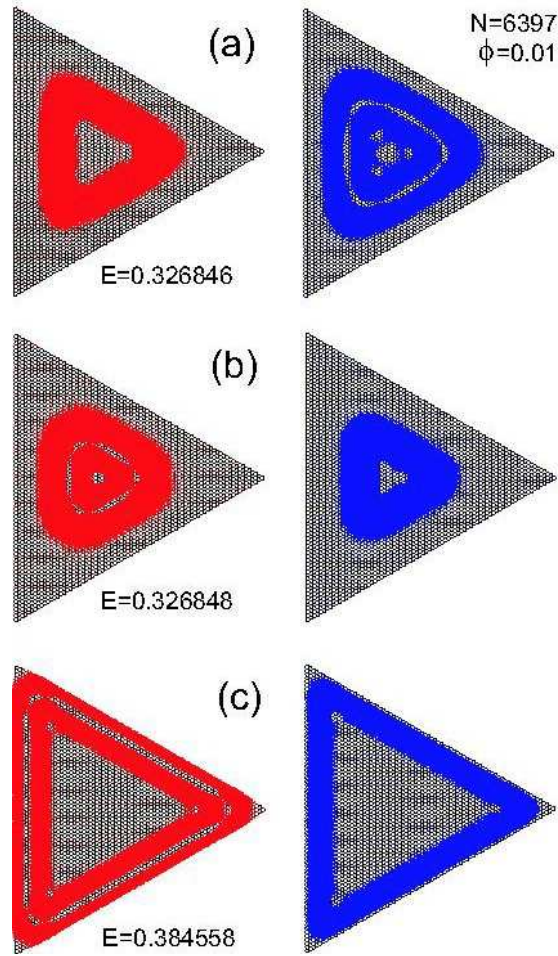}
\caption{%%%%
Examples of TB electron densities at high magnetic field ($\phi=0.01$) for the A 
(left) and B (right) sublattices associated with: (a)$+$(b) double-bulk states in the $n=1$ 
Landau level. (c) double-edge states situated on the rising-in-energy branch between the $n=1$ 
and $n=2$ Landau levels.
Energies in units of the hopping coupling parameter $t$. 
}
\label{abtbLLn1}% must come after the caption
\end{figure}
%*********************** end figure 7 **************

In Fig.\ \ref{enetb}, we display the TB energies for the triangular flake with $\phi=0.01$,
$\phi=eSB/(hc)$ being the dimensionless magnetic flux through an hexagonal unit of the 
two-dimensional graphene lattice; $S$ is the area enclosed by the hexagon and $B$ is the 
magnetic field. The value $\phi=0.01$ corresponds to a magnetic field sufficiently high so that
Landau levels have been formed, but at the same time low enough so that Hofstadter-butterfly
\cite{hofs76} effects (due to the periodicity of the lattice) have not developed. \cite{peet08} 
The TB-energy curve in Fig.\ \ref{enetb} exhibits a well defined trend, i.e., several 
almost-flat horizontal segments are connected via fast-varying and rising-in-energy branches. 
The flat segments correspond to the Landau levels with energy 
$E/t \propto {\text{sign}}(n) \sqrt{|n|}$, $n=0, \pm 1, \pm 2,\ldots$. 

A close inspection of the properties and behavior of the states associated with the TB energies
in the $n=0$ level reveals that this level contains two
different bands: (i) a nondispersive one with energies close to the available machine 
precision $(E/t \sim 10^{-14})$  [that correspond to the $\varepsilon=0$ solutions 
of the continuous model; see Sec.\ \ref{secdw}] and (ii) a quasidegenerate dispersive band 
with energies that are still very small (starting at $E/t \sim 10^{-6}$) but increase 
gradually and then merge with the rising-in-energy branch; this band corresponds [see the 
transcendental equation (\ref{eqtran})] to the $\varepsilon \gtrsim 0$ solutions of the 
continuous model in the flat region of the lowest curve in Fig.\ \ref{ene}(a). 
The nondispersive zero states coincide with the midgap surface states under field-free 
conditions and they have been studied extensively; 
\cite{fuji96,dres96,waka99,son06,brfr06.1,wimm10} 
these states are not influenced by the magnetic field and will not be discussed
any further here. The close-to-zero dispersive states are formed as a consequence of the 
external magnetic field, and they will be the primary focus of this paper.  

%*********************** begin figure 8 **************
\begin{figure}[t]
\centering\includegraphics[width=7.5cm]{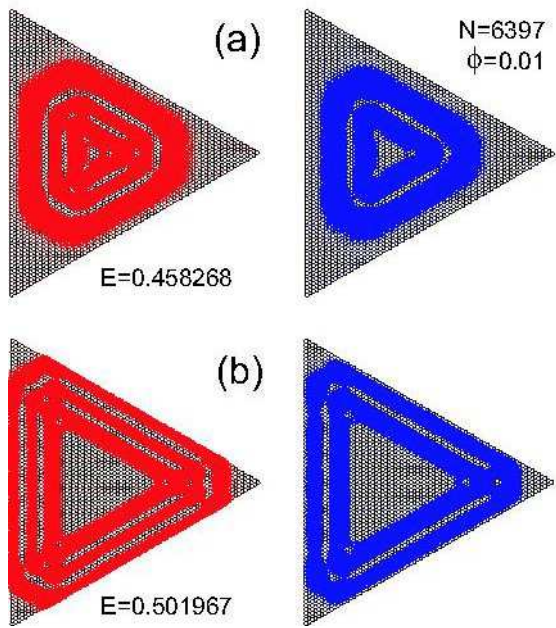}
\caption{%%%%
Examples of TB electron densities at high magnetic field ($\phi=0.01$) for the A 
(left) and B (right) sublattices of a triangular flake with $N=6397$ carbon atoms and zigzag 
edges. (a) double-bulk states in the $n=2$ Landau level. (b) double-edge states situated on the
rising-in-energy branch between the $n=2$ and $n=3$ Landau levels. 
Energies in units of the hopping coupling parameter $t$.
}
\label{abtbLLn2}% must come after the caption
\end{figure}
%*********************** end figure 8 **************

In Fig.\ \ref{abtbLLL}, we display the TB electron densities (specifically the square root of 
the densities \cite{note2}) for states associated with the LLL in the case of a triangular 
graphene flake under the same magnetic field $\phi=0.01$. The electron densities 
of the A (left, online red) and B (right, online blue) 
sublattices are plotted separately. Figs.\ \ref{abtbLLL}(a) and \ref{abtbLLL}(b) correspond
to TB states with energies ($E/t=0.1699 \times 10^{-7}$ and $E/t=0.1475 \times 10^{-3}$,
respectively) that lie well inside the flat segment of the LLL ($n=0$) energy curve in
Fig.\ \ref{enetb}. It is apparent that these states are of a mixed bulk-edge character, with
the B-sublattice component being bulk-like and the A-sublattice component being edge-like.
They are analogous to the mixed bulk-edge LLL states described in Sec.\ \ref{secdw} within the
continuous relativistic Dirac-Weyl-equation framework. In particular, the spatial profile of 
the A-sublattice component of the TB densities in Figs.\ \ref{abtbLLL}(a) and \ref{abtbLLL}(b) 
agrees well with the surface-state Dirac-spinor component in Eq.\ (\ref{dm12}); see also
Fig.\ \ref{chibaLLL}, bottom frame. Moreover, the profiles of the TB densities for the B 
sublattices in these figures exhibit the qualitative behavior of a $D_\nu[\sqrt{2}(x-x_c)]$ 
function with $\nu \gtrsim 0$ described in Sec.\ \ref{secdw}; see also Fig.\ \ref{chibaLLL}, 
top frame. Note that a lower TB energy [case of Fig.\ \ref{abtbLLL}(a)] corresponds to a 
continuous $D_\nu[\sqrt{2}(x-x_c)]$ state with a larger centroid $x_c > 0$, farther away from
the physical edge. For states with higher TB energies, lying on the rising-in-energy branch,
the B-sublattice component moves towards the physical edge and transforms into an edge state, 
as illustrated by the double-edge TB state in Fig.\ \ref{abtbLLL}(c) (with energy 
$E/t=0.3033$).

In Figs.\ \ref{abtbLLn1}(a) and \ref{abtbLLn1}(b), we portray TB densities for two states 
associated with the second Landau level (with index $n=1$ at an energy $E/t \approx 0.32685$).
Fig.\ \ref{abtbLLn1}(a) corresponds to a Dirac spinor in the $K$ valley having an A component
consisting of a $D_0(\xi)$ state (no nodes inside the graphene flake) and a B component
consisting of a $D_1(\xi)$ state (a single node inside the graphene flake). Fig.\ 
\ref{abtbLLn1}(b) portrays a similar TB state in the $K^\prime$ valley, since the A and B 
sublattices correspond to the continuous functions $D_1(\xi)$ and $D_0(\xi)$, respectively, 
which is the opposite from the $K$-valley case in Fig.\ \ref{abtbLLn1}(a); see Sec. 
\ref{seckpr}. Fig.\ \ref{abtbLLn1}(c) portrays a double-edge state with energy $E/t=0.384558$ 
(lying in Fig.\ \ref{enetb} on the rising-in-energy branch between the $n=1$ and $n=2$ Landau 
levels). It is apparent that this double-edge TB state is associated with the $K^\prime$ 
valley and has been evolved out of the double-bulk state in Fig.\ \ref{abtbLLn1}(b); note
the preservation of the single node (no node) topology in the A-sublattice (B-sublattice) 
electron-density component.

In Fig.\ \ref{abtbLLn2}(a), we portray TB densities for a double-bulk state associated with 
the third Landau level (with index $n=2$ at an energy $E/t \approx 0.458268$; see Fig.\ 
\ref{enetb}). The TB densities in Fig.\ \ref{abtbLLn2}(a) correspond to a continuous Dirac 
spinor in the $K^\prime$ valley having an A component consisting of a 
$D_2(\xi)$ state (two nodes inside the graphene flake) and a B component consisting of a 
$D_1(\xi)$ state (a single node inside the graphene flake); see Sec. \ref{seckpr} and Fig.\
Fig.\ \ref{chibaLL2} [but with B replaced by A$^\prime$ and A replaced by B$^\prime$
(online: blue $\leftrightarrow$ red)]. Fig.\ \ref{abtbLLn2}(b) portrays a 
double-edge state with energy $E/t=0.501967$ (lying in Fig.\ \ref{enetb} on the 
rising-in-energy branch between the $n=2$ and $n=3$ Landau levels). It is apparent that this 
double-edge TB state is associated also with the $K^\prime$ valley and has been evolved out of 
the double-bulk state in Fig.\ \ref{abtbLLn2}(a); note the preservation of the two-node 
(single-node) topology in the A-sublattice (B-sublattice) electron-density component. Note
the similarities with the double-edge continuous Dirac spinor within the physical semi-infinite 
graphene plane portrayed in Fig.\ \ref{chibaLL22}.

%*********************** begin figure 9 **************
\begin{figure}[t]
\centering\includegraphics[width=7.5cm]{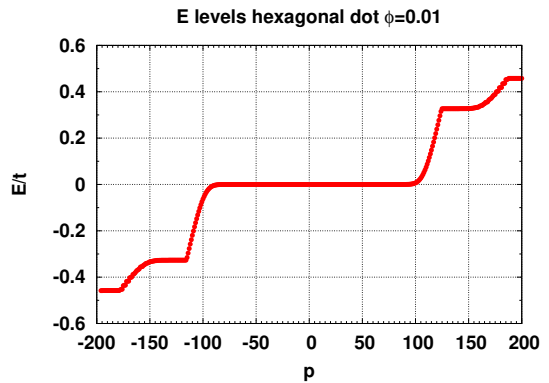}
\caption{%%%%
The TB energies at high magnetic field ($\phi=0.01$) for an hexagonal graphene flake with 
zigzag edges comprising $N=6144$ carbon atoms. The $n=0$, $\pm 1$, and $\pm 2$ Landau levels 
correspond to the flat segments of the curve. The double-edge states correspond to the 
rising-in-energy branches between the flat segments. The integer index $p$ counts the TB states
(negative $p$ values correspond to negative energies).
}
\label{enetbhex}% must come after the caption
\end{figure}
%*********************** end figure 9 **************

\subsection{Hexagonal graphene flakes}
\label{sechex}

In Sec. \ref{sectri}, we studied the nature of the TB states in the case of a trigonal
flake with zigzag edge terminations. A characteristic property of triangular flakes is that 
the same sublattice participates in the edge terminations of two adjacent polygonal sides 
(forming an angle of 60$^\circ$). In this section, we study hexagonal graphene flakes 
\cite{peet08} with zigzag edge terminations, which is a more complicated case. This is due to 
the fact that the sublattices A and B alternate in providing the edge termination of adjacent 
polygonal sides (having an angle of 120$^\circ$).

In Fig.\ \ref{enetbhex}, we display the TB energies for an hexagonal flake with $\phi=0.01$ 
(corresponding to a high magnetic field where Landau levels have been formed). As was the case
with the trigonal TB energies in Fig.\ \ref{enetb}, the TB-energy curve in Fig.\ \ref{enetbhex}
exhibits also a well defined trend, i.e., several almost-flat horizontal segments which are 
connected via fast-varying (rising-in-energy) branches. 
The flat segments correspond to the Landau levels with energies 
$E \propto {\text{sign}}(n) \sqrt{|n|}$, $n=0, \pm 1, \pm 2,\ldots$. 

Furthermore, as was also the case with the trigonal flakes, a close inspection of the 
properties and behavior of the states associated with the TB energies in the $n=0$ level
in Fig.\ \ref{enetbhex} reveals that this level contains two
different bands: (i) a nondispersive one with energies close to the available machine 
precision $(E/t \sim 10^{-14})$ [that correspond to the $\varepsilon=0$ solutions 
of the continuous model; see Sec.\ \ref{secdw}] and (ii) a quasidegenerate dispersive band 
with energies that are still very small (starting at $E/t \sim 10^{-6}$) but increase 
gradually and then merge with the rising energy branch; this band corresponds [see the 
transcendental equation (\ref{eqtran})] to the $\varepsilon \gtrsim 0$ solutions of the 
continuous model in the flat region of the lowest curve (i.e., in the LLL) in Fig.\ 
\ref{ene}(a). Here, we study these dispersive states in the LLL that are formed due to the 
presence of the magnetic field.

%*********************** begin figure 10 **************
\begin{figure}[t]
\centering\includegraphics[width=7.5cm]{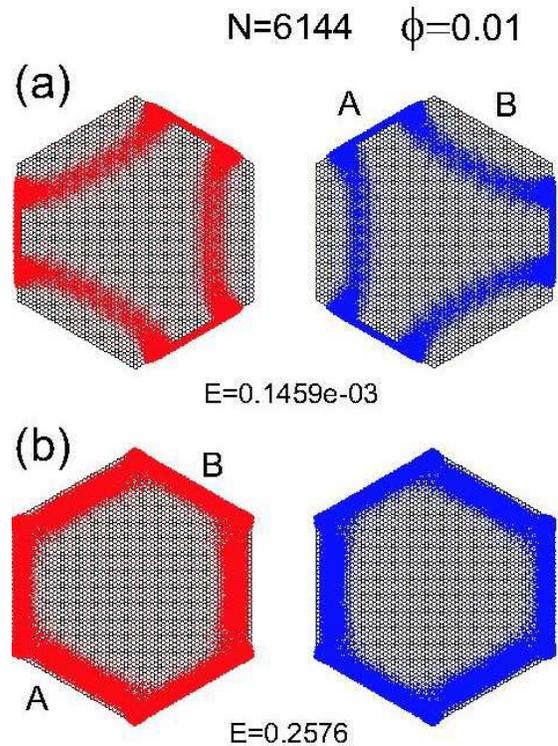}
\caption{%%%%
Examples of TB electron densities at high magnetic field ($\phi=0.01$) for the A 
(left) and B (right) sublattices of an hexagonal flake with $N=6144$ carbon atoms and zigzag 
edges. (a) a mixed bulk-edge state in the $n=0$ Landau level. (b) a double-edge state with 
energy situated on the rising-in-energy branch between the $n=0$ and $n=1$ Landau levels. 
The labels A and B indicate the (alternating) sublattice-type edge termination along the sides 
forming the physical boundary of the hexagon.  
Energies in units of the hopping coupling parameter $t$.
}
\label{abtbLLLhex}% must come after the caption
\end{figure}
%*********************** end figure 10 **************

In Fig.\ \ref{abtbLLLhex}, we display the TB electron densities 
(specifically the square root of the densities \cite{note2}) 
for states associated with the LLL in the case of an hexagonal graphene flake 
under the same magnetic field $\phi=0.01$. The electron densities of the A (left, 
online red) and B (right, online blue) sublattices are plotted separately. Fig.\ 
\ref{abtbLLLhex}(a) corresponds to a TB state with energy ($E/t=0.1459 \times 10^{-3}$) 
that lies well inside the flat segment of the LLL ($n=0$) energy curve in Fig.\ \ref{enetbhex}.
This state is of a mixed bulk-edge character, exhibiting, however, a more complex profile
compared to the corresponding mixed bulk-edge states for the trigonal flake in Figs.\ 
\ref{abtbLLL}(a) and \ref{abtbLLL}(b). This is due to the alternation of the A and B 
sublattices along the polygonal sides forming the edge of the hexagonal flake. 
In particular, focusing on a side with a B-sublattice termination (e.g., the one at the 
upper-right corner), one sees that the A-sublattice density (left, online red) exhibits an 
edge-state behavior, while the B-sublattice density (right, online blue) exhibits a bulk-state 
profile. Focusing on a side with an A-sublattice termination (e.g., the one at the lower-left
corner), one sees the opposite, i.e., the A-sublattice density (left, online red) exhibits a 
bulk-state behavior, while the B-sublattice density (right, online blue) exhibits an edge-state
profile. In the continuous relativistic Dirac-Weyl model (Sec. \ref{secdw}), the former case is 
associated with spinor components in the $K$ valley (permitting the vanishing of $\chi_B$ on 
the edge; see Sec. \ref{seck}), while the latter is associated with spinor components in the
$K^\prime$ valley (permitting the vanishing of $\chi_A^\prime$ on the edge; see Sec. 
\ref{seckpr}).

%*********************** begin figure 11 **************
\begin{figure}[t]
\centering\includegraphics[width=7.5cm]{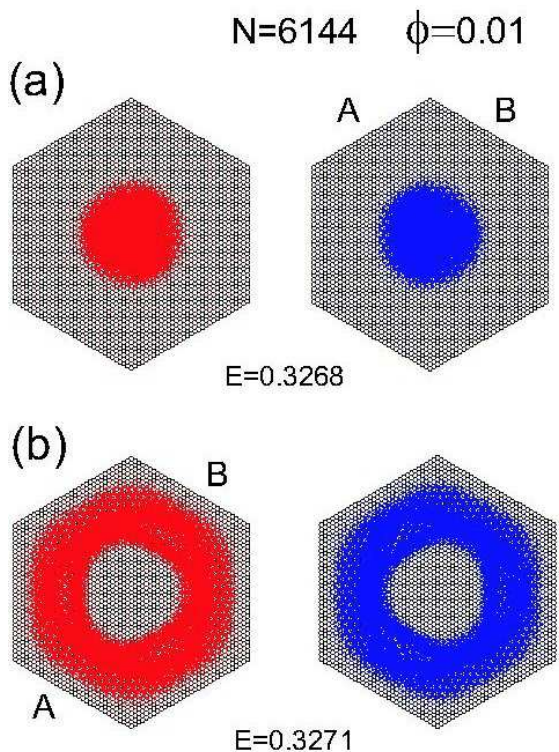}
\caption{%%%%
Examples of TB electron densities at high magnetic field ($\phi=0.01$) for the A 
(left) and B (right) sublattices of an hexagonal flake with $N=6144$ carbon atoms and zigzag 
edges. Both rows [(a) and (b)] portray double-bulk spinors in the $n=1$ Landau level. 
The labels A and B indicate the (alternating) sublattice-type edge termination along the sides
forming the physical boundary of the hexagon.
Energies in units of the hopping coupling parameter $t$.
}
\label{abtbLLn1hex}% must come after the caption
\end{figure}
%*********************** end figure 11 **************

Fig.\ \ref{abtbLLLhex}(b) displays the TB densities for states with a higher TB energy, 
$E/t=0.2576$, lying on the rising-in-energy branch between the $n=0$ and $n=1$ Landau levels
in Fig.\ \ref{enetbhex}. This state represents a double-edged one and 
it can be interpreted as having continuously evolved form the state portrayed in
Fig.\ \ref{abtbLLLhex}(a), with the bulk components having been pushed against the edges.

%*********************** begin figure 12 **************
\begin{figure}[t]
\centering\includegraphics[width=7.5cm]{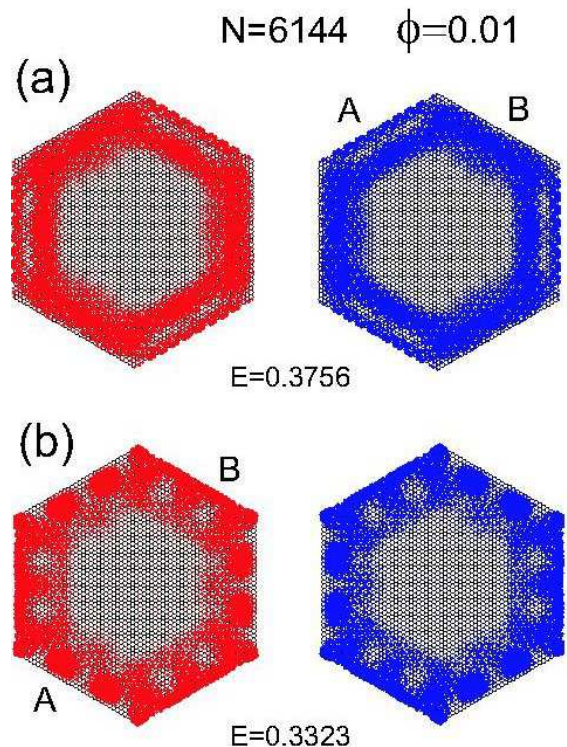}
\caption{%%%%
Examples of TB electron densities at high magnetic field ($\phi=0.01$) for the A 
(left) and B (right) sublattices of an hexagonal flake with $N=6144$ carbon atoms and zigzag 
edges. Both rows [(a) and (b)] portray double-edge spinors with energies in the 
rising-in-energy branch between the $n=1$ and $n=2$ Landau level. 
The labels A and B indicate the (alternating) sublattice-type edge termination along the sides
forming the physical boundary of the hexagon.
Energies in units of the hopping coupling parameter $t$.
}
\label{abtbLLn1edgehex}% must come after the caption
\end{figure}
%*********************** end figure 12 **************

Fig.\ \ref{abtbLLn1hex} portrays TB densities for two double-bulk states associated with 
the second Landau level of the hexagonal flake (with index $n=1$ at an energy 
$E/t \approx 0.3270$; see Fig.\ \ref{enetbhex}). The sublattice densities in Fig.\ 
\ref{abtbLLn1hex}(a) correspond to a TB state with the lowest energy ($E/t \approx 0.3268$) in 
this Landau level, and they are concentrated around the center of the flake, far away from the 
edges. In addition, neither one of them (A-sublattice density or B-sublattice density) exhibits
any nodes, which seems paradoxical at a first glance for a state belonging to the first Landau 
level. The explanation can be found in that the edge of the flake has very little influence in 
this case, which thus corresponds to a nodeless $l=0$ Dirac-Weyl state of the $n=1$ LL in a 
circular graphene dot with zigzag termination; see Eqs.\ (A.2) and (A.4) in Ref.\ 
\onlinecite{yrl10}.  

Fig.\ \ref{abtbLLn1hex}(b) portrays TB densities for another double-bulk state with energy 
$E/t=0.3271$ belonging to the second ($n=1$) Landau level of the hexagonal flake; see Fig.\ 
\ref{enetbhex}. The density profiles in Fig.\ \ref{abtbLLn1hex}(b) exhibit a zero-node and
a one-node structure in analogy with the $D_0(\xi)$ and $D_1(\xi)$ functions describing the 
$n=1$ Landau level in the continuous Dirac-Weyl model (see Sec. \ref{secdw}). However, in 
contrast to the single-edge semiinfinite graphene plane in Sec. \ref{secdw}, in Fig.\ 
\ref{abtbLLn1hex}(b) the zero-node/one-node topology is present in both sublattices (in both 
the left and right panels) due to the alternation of the edge termination along the hexagon's 
sides between the A and B sublattices.

Fig.\ \ref{abtbLLn1edgehex} portrays TB densities for two double-edge states whose energies
lie on the rising-in-energy branch between the $n=1$ and $n=2$ Landau levels; see Fig.\
\ref{enetbhex}. The sublattice densities in Fig.\ \ref{abtbLLn1edgehex}(a) correspond to a TB 
state with energy $E/t=0.3756$. It is apparent that they can be viewed as having evolved out of
the densities in Fig.\ \ref{abtbLLn1hex}(b), with the centroids of the bulk densities having
been pushed against the hexagonal edges. Remarkably, at the same time, the zero-node/one-node
alternating nodal topology is preserved in the double-edge state in Fig.\ 
\ref{abtbLLn1edgehex}(a) in complete ananlogy with the double-bulk state in Fig.\ 
\ref{abtbLLn1hex}(b).   

The sublattice densities in Fig.\ \ref{abtbLLn1edgehex}(b) correspond to a TB
state with energy $E/t=0.3323$; they also represent a double-edge state
like the one in Fig.\ \ref{abtbLLn1edgehex}(a). However, a new characteristic [compared to the
double-edge state in Fig.\ \ref{abtbLLn1edgehex}(a)] is the appearance of an additional nodal 
pattern resulting from wave-function quantization along (and parallel to) the hexagon's sides. 
This additional pattern is superimposed upon the perpendicular-to-the-edge 
zero-node/one-node pattern (the latter being present already in the continuous Dirac-Weyl model
of a semiinfinite graphene plane studied in Sec. \ref{secdw}). We note that such combined
parallel-to and perpendicular-to-the-edge nodal patterns (not shown) have been found by us 
also in several instances of TB states in the case of a trigonal graphene flake in Sec. 
\ref{sectri}.
  
%*********************** begin figure 13 **************
\begin{figure}[t]
\centering\includegraphics[width=7.5cm]{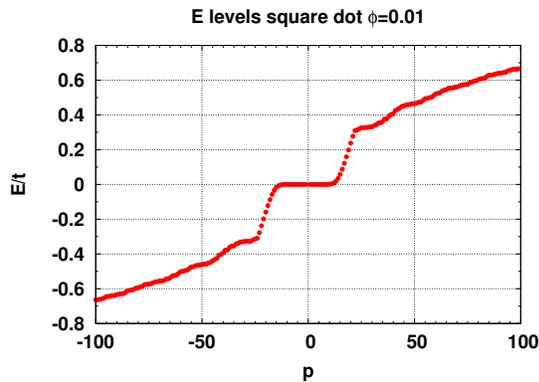}
\caption{%%%%
The TB energies at high magnetic field ($\phi=0.01$) for a square graphene flake with both 
zigzag and armchair edges comprising $N=2074$ carbon atoms. The lowest ($n=0$) Landau level 
corresponds to the flat segment of the curve at $E \approx 0$. The integer index $p$ counts the
TB states (negative $p$ values correspond to negative energies).
}
\label{enetbrec}% must come after the caption
\end{figure}
%*********************** end figure 13 ****************

%*********************** begin figure 14 **************
\begin{figure}[t]
\centering\includegraphics[width=7.5cm]{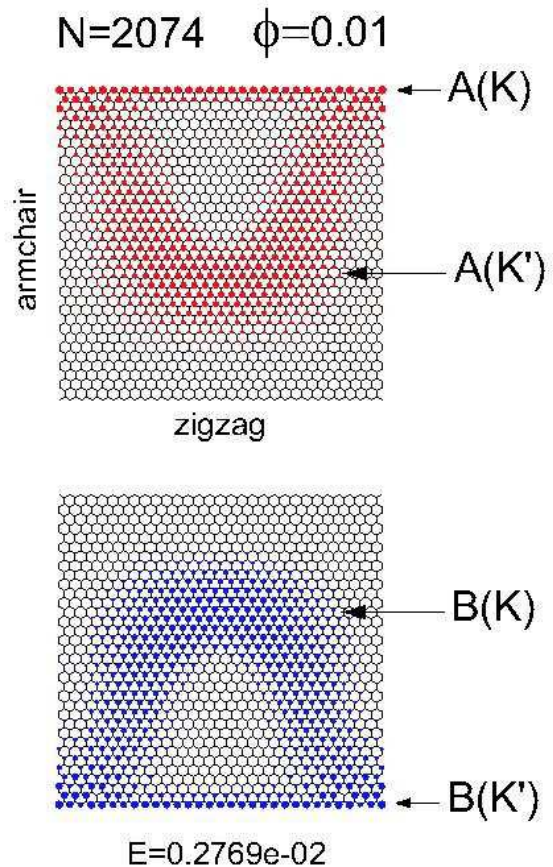}
\caption{%%%%
Example of TB electron densities at high magnetic field ($\phi=0.01$) for the A (online 
red; top frame) and B (online blue; bottom frame) sublattices of a square flake with 
$N=2074$ carbon atoms and both zigzag and armchair edges. A state representing a mixed 
bulk-edge state in the LLL is displayed. The density contributions on each sublattice associated
with a given valley are also marked. Energies in units of the hopping coupling parameter $t$.
}
\label{abtbLLLrec}% must come after the caption
\end{figure}
%*********************** end figure 14 **************

\subsection{Square graphene flakes}
\label{secsqu}

In this section, we study the more complicated case of a square graphene flake which 
necessarily has mixed armchair and zigzag edge terminations (each type of termination developing
on opposite sides of the square). In Fig.\ \ref{enetbrec}, we display the TB energies for such 
a square flake with $\phi=0.01$ as in the previous studied cases of trigonal (Sec. 
\ref{sectri}) and hexagonal flakes (Sec. \ref{sechex}). Compared to the energy curves in the 
previous cases [trigonal (see Fig.\ \ref{enetb}) and hexagonal (see Fig.\ \ref{enetbhex})],
the TB energies in Fig.\ \ref{enetbrec} exhibit higher Landau levels (horizontal segments in 
Fig.\ \ref{enetbrec}) with $n \ge 1$ that are not as well formed; this may be due to the 
smaller number of carbon atoms in the square flake $(N=2074)$. The LLL ($n=0$) level, however, 
is well formed, and this is sufficient for our purposes here, namely to investigate whether the
mixed LLL bulk-edge states maintain in the presence of edge segments with armchair termination.

Indeed the TB sublattice densities for the LLL state (with energy $E/t=0.2769 \times 10^{-2}$) 
portrayed in Fig.\ \ref{abtbLLLrec} show that the mixed LLL bulk-edge behavior maintains also 
in the case of a square flake. Naturally, due to the coupling between the $K$ and $K'$ valleys
induced by the presence of the armchair terminations, each sublattice [A (online red) and 
B (online blue)] exhibits now both edge and bulk density contributions, which however 
correspond to different valleys as explicitly marked in Fig.\ \ref{abtbLLLrec}.

Mixed bulk-edge states of a square graphene flake in a perpendicular magnetic field were also 
reported in a recent study. \cite{eyang10} In this study the appearance of such mixed states 
with significant weight at the zigzag edges were attributed to the coupling between the $K$ and
$K^\prime$ valleys due to the armchair edges (see in particular Sec. IV in Ref.\ 
\onlinecite{eyang10}). This interpretation differs from the conclusion presented in our paper
where the occurrence of mixed bulk-edge LLL states is shown to originate solely from the
zigzag edge termination.

\section{Summary and Discussion}
\label{secsum}

The properties of single-electron states in graphene flakes with zigzag edge termination under
high magnetic fields (in the regime of Landau-level formation) were investigated using 
tight-binding calculations. A systematic interpretation of their character (bulk-like versus
edge-like) was achieved via a comparison of the tight-binding electron densities with analytic
expressions (based on parabolic cylinder functions) for the relativistic Dirac-Weyl spinors in 
the case of a semi-infinite graphene plane. A variery of graphene flakes was considered,
namely, trigonal, hexagonal, and square ones.

The higher Landau levels were found to comprise exclusively electrons of bulk-type character
(for both sublattices). Furthermore, electrons with energies on the rising-in-energy branches 
(connecting the Landau levels) are described by edge-type states reminiscent of those familiar 
from the theory of the integer quantum Hall effect for nonrelativistic electrons. 

In contrast, in all cases studied the lowest, ($n=0$) Landau level contained relativistic Dirac 
electrons of a mixed bulk-edge character without an analog in the nonrelativistic case. Most
importantly, it was shown that such mixed bulk-edge states maintain also in the case of a 
square flake with combined zigzag and armchair edge terminations. 

The presence of mixed bulk-edge LLL states in graphene samples with realistic shapes points at
significant implications concerning the many-body correlated FQHE excitations.
We recall that Ref.\ \onlinecite{yrl10} studied the many-body correlated FQHE excitations in 
the LLL in the simplified case of a circular graphene flake with zigzag edge termination,
and it found that the two-body Coulomb-interaction matrix elements are given as a sum of four
terms
\begin{equation}
\frac{1}{4}( \langle \tilde{b}_1 \tilde{b}_2 | \tilde{b}_3 \tilde{b}_4 \rangle +
\langle \tilde{e}_1 \tilde{e}_2 | \tilde{e}_3 \tilde{e}_4 \rangle +
\langle \tilde{b}_1 \tilde{e}_2 | \tilde{b}_3 \tilde{e}_4 \rangle +
\langle \tilde{e}_1 \tilde{b}_2 | \tilde{e}_3 \tilde{b}_4 \rangle),
\label{cmeexp}
\end{equation}
where $|\tilde{b} \rangle$ and $| \tilde{e} \rangle$ denote the bulk and edge components of
the mixed LLL state; note that, due to the equal weights $(50\%-50\%)$ of the bulk-like and 
edge-like components, a prefactor of 1/4 appears in front of each term in Eq.\ (\ref{cmeexp}).
As a consequence of Eq.\ (\ref{cmeexp}) and of the 1/4 prefactor, a sizable attenuation of 
the many-body correlated FQHE excitations in the LLL (associated with the 
$\langle \tilde{b}_1 \tilde{b}_2 | \tilde{b}_3 \tilde{b}_4 \rangle/4$ term reflecting the
depletion of the bulk component) was found in the 
simplified case of a circular graphene flake. Furthermore, it was shown \cite{yrl10} that the 
insulating behavior at the Dirac neutrality point under high $B$ (experimentally observed 
\cite{andr09,kim09} in graphene samples along with the $1/3$ FQHE) is associated with the 
Coulombic repulsion due to the accumulation of charge at the edges [related to the 
$\langle \tilde{e}_1 \tilde{e}_2 | \tilde{e}_3 \tilde{e}_4 \rangle/4$ and the remaining 
two cross terms in the Coulomb-interaction matrix elements given in Eq.\ (\ref{cmeexp})].

The current study shows that the appearance of mixed bulk-edge LLL states is a property of
the presence of segments in the graphene-sample boundary having a zigzag edge termination,
independent of the precise shape of the graphene sample. This finding suggests that  
the results of Ref.\ \onlinecite{yrl10} concerning the many-body correlated FQHE excitations in
the LLL can be generalized to graphene samples with more realistic shapes.

\begin{acknowledgments}
This work was supported by the Office of Basic Energy Sciences of the US D.O.E. 
under contract FG05-86ER45234.
\end{acknowledgments}


\begin{thebibliography}{999}
\bibitem{note1}
The ceiling function is defined as $\lceil \nu \rceil=n+1$, if $n< \nu \leq n+1$ with
$n$ being an integer.
\bibitem{note2}
A scaling factor is used in order to achieve the best legibility of the figures.
\end{thebibliography}
\end{document}